\documentclass[usenatbib,twocolumn,twocolappendix,tighten]{aastex63}

\usepackage{graphicx} \usepackage{amssymb} \usepackage{amsmath}
\usepackage{epstopdf} \usepackage{aas_macros} \usepackage{color}

\newcommand{\cpm}{{\small CUBEP$^3$M}}

\newcommand{\cube}{{\small CUBE}}

\newcommand{\camb}{{\small CAMB}}

\newcommand{\ev}{{\rm eV}}

\newcommand{\hmpc}{$h\ ${\rm Mpc}$^{-1}$}

\newcommand{\mpch}{$h^{-1}${\rm Mpc}}

\newcommand{\kms}{{\rm km s}$^{-1}$}

\newcommand{\cnb}{{C$\nu$B}}

\newcommand{\ft}{{\rm FT}}

\begin{document}
\title{Simulating the Cosmic Neutrino Background using Collisionless
  Hydrodynamics}

\author[0000-0001-6039-9058]{Derek Inman} \email{derek.inman@nyu.edu}
\affiliation{Center for Cosmology and Particle Physics, Department of
  Physics, New York University, 726 Broadway, New York, NY, 10003,
  USA}
  
\author[0000-0001-5277-4882]{Hao-Ran Yu} \email{haoran@xmu.edu.cn}
\affiliation{Department of Astronomy, Xiamen University, Xiamen,
  Fujian, 361005, China}

\begin{abstract}
  The cosmic neutrino background is an important component of the
  Universe that is difficult to include in cosmological simulations
  due to the extremely large velocity dispersion of neutrino
  particles.  We develop a new approach to simulate cosmic neutrinos
  that decomposes the Fermi-Dirac phase space into shells of constant
  speed and then evolves those shells using hydrodynamic equations.
  These collisionless hydrodynamic equations are chosen to match
  linear theory, free particle evolution and allow for superposition.
  We implement this method into the information-optimized cosmological
  $N$-body code \cube{} and demonstrate that neutrino perturbations
  can be accurately resolved to at least $k\sim1$ \hmpc.  This
  technique allows for neutrino memory requirements to be decreased by
  up to $\sim 10^3$ compared to traditional $N$-body methods.
\end{abstract}

\section{Introduction}
The cosmic neutrino background (\cnb) has been robustly detected by
observations of the cosmic microwave background perturbations (CMB)
\citep{bib:planck2018}.  At $z\sim1100$ when the primary CMB
perturbations are set the neutrinos are highly relativistic; however,
observations of neutrino oscillations \citep{bib:deSalas2017} suggest
that at least one species of neutrino has mass $m_\nu \gtrsim 0.05$
\ev{} and so should be non-relativistic today.  Measuring the
properties of the non-relativistic \cnb{} has become one of the chief
goals of upcoming cosmological experiments.  Of principle importance
is a measurement of the total neutrino mass $M_\nu = \sum m_\nu$
which, in conjunction with oscillation experiments, may also resolve
the neutrino hierarchy and whether any neutrinos are massless.
Currently cosmological constraints
(e.g.~\citet{bib:planck2018,bib:PalanqueDelabrouille}) of $M_\nu$ are
significantly better than terrestrial ones \citep{bib:Aker2019}.

The \cnb{} is distinct in that it becomes massive but still remains
quite hot and so does not cluster nearly as much on small scales as
the dominant cold dark matter (CDM).  The principle effect of this
lack of clustering is a modulation of the matter power spectrum in a
way that is sensitive to the neutrino energy density $\Omega_\nu$.  If
standard cosmology holds, where neutrinos are the only hot species and
have the standard decoupling temperature
$T_\nu = (4/11)^{1/3} T_{\rm CMB}$, then
$\Omega_\nu h^2 = M_\nu/93.14$ \citep{bib:Mangano2006}.  This
technique for determining $M_\nu$ is a substantial challenge both
because the modulation is small and because we principally measure not
the underlying matter field but rather biased tracers of it such as
galaxies.  If $M_\nu$ is minimal, forecasts for DESI, EUCLID and
CMB-S4 suggest that near future detections of the \cnb{} using the
matter power spectrum will be at most around $3\sigma$
\citep{bib:Brinkmann2019}.  Finding other observables sensitive to
neutrinos is therefore critical to improve the robustness of the
non-relativistic \cnb{} detection.  Alternatives that have been
suggested include void statistics
\citep{bib:Massara2015,bib:Banerjee2016,bib:Kreisch2019}, the relative
velocity effect
\citep{bib:Zhu2014,bib:Zhu2016,bib:Inman2015,bib:Inman2017b,bib:Zhu2019},
scale-dependent halo bias
\citep{bib:loverde2014,bib:Chiang2018,bib:Chiang2019,bib:Banerjee2019},
differential neutrino condensation \citep{bib:Yu2017}, and via galaxy
spins \citep{bib:Yu2019}.

It is therefore critical to accurately model the effects of neutrinos
on large scale structure.  The \cnb{} has been included in simulations
in a number of ways.  The most straightforward methods invoke linear
theory, either interpolating between precomputed results
\citep{bib:Brandbyge2009} or integrating alongside the simulation
\citep{bib:AliHaimoud2013}.  These methods are quite fast and lead to
the correct modulation of the matter power spectrum, but do not
accurately resolve neutrino dynamics.  The standard method to obtain
correct neutrino behaviour is to use N-body particles which have large
thermal velocities drawn from the Fermi-Dirac distribution,
e.g.~\citet{bib:Brandbyge2008,bib:Viel2010,bib:Zennaro2017}.  The
chief downside of this style of simulation is that the random
velocities are so large that they introduce significant randomness
into the neutrino density field that substantially erases the true
perturbations on small scales.  The effects of this Poisson noise can
be suppressed through the use of many particles
\citep{bib:Emberson2017} or through the use of hybrid methods where
part of the evolution is linear
\citep{bib:Brandbyge2010,bib:Inman2015,bib:Bird2018}.

There are two methods employed in the literature that are Poisson
noise free.  The first is to use neutrino particles, but introduce
regularity in the thermal velocities such that random clustering does
not occur \citep{bib:Banerjee2018}.  In practice, this involves
utilizing many particles per grid cell, with each cell having the same
sampling of random velocities.  The second is to instead solve the
Boltzmann moment equations on a grid, which has no Poisson noise by
construction.  The difficulty here is that the Boltzmann moment
equations are an infinite hierarchy of equations with no known
truncation scheme consistent with the neutrino distribution.  One
approach is to close the hierarchy by utilizing linear theory for the
stress terms in the momentum equation \citep{bib:Dakin2017} which
works well provided that the pressure is always correlated with the
density field.  The other method is to use N-body particles for
closure by using them to estimate the stress terms (instead of the
density) \citep{bib:Banerjee2016}.

\begin{figure}
  \begin{center}
    \includegraphics[width=0.475\textwidth]{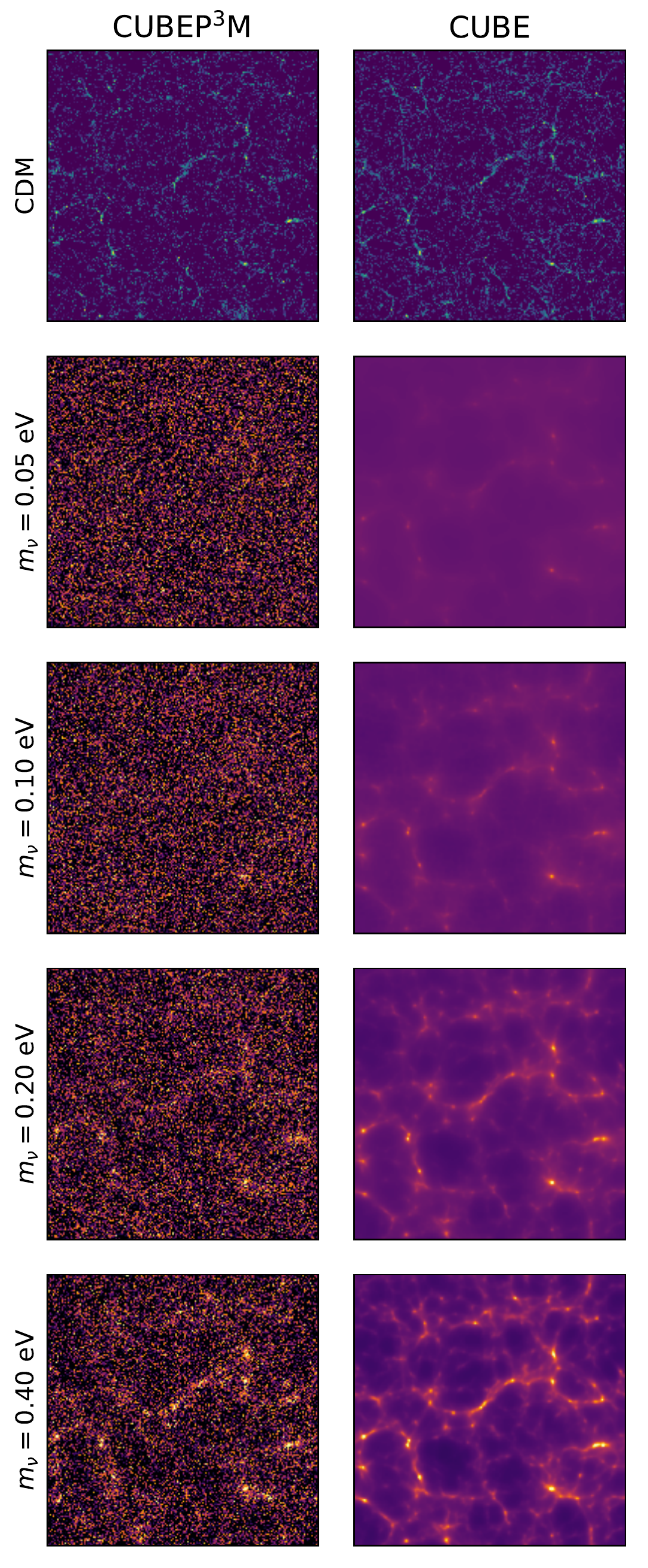}
    \caption{Slices of CDM (top row) and \cnb{} (bottom four rows)
      density fields computed using \cpm{} (left) and \cube{} (right).
      The slices are 350 \mpch{} in width and 0.7 \mpch{} in depth.
      The hydrodynamic method used in \cube{} is free from Poisson
      noise.}
    \label{fig:slices}
  \end{center}
\end{figure}

In this paper we also solve the Boltzmann moment equations; however we
do so by using a closure scheme that does not rely on either external
transfer functions or N-body particles.  We do this by decomposing the
neutrino phase space into shells of uniform speed, for which a
straightforward linear closure scheme exists.  It is conceptually
quite similar to the multiflow description of
\citet{bib:Dupuy2014,bib:Dupuy2015}, although we consider particles
with the same speed rather than the same velocity.  We show slices
comparing our method and standard N-body in Fig.~\ref{fig:slices}.
The lack of Poisson noise is immediately evident. In Section
\ref{sec:theory} we motivate the closure scheme for the velocity
shells.  In Section \ref{sec:methods} we discuss our numerical
implementation of the method into the \cube{} code \citep{bib:Yu2018}.
We then show results of these simulations, including comparisons to
standard N-body methods, in Section \ref{sec:results}.  We discuss
future optimizations and extensions in Sections \ref{sec:discussion}
and \ref{sec:conclusion}.

\section{Theory}
\label{sec:theory}

\subsection{Vlasov Equation and Boltzmann Moment Equations}
The phase space, $f(t,x^i,v^i)$, of collisionless particles is
described by the Vlasov equation:
\begin{align}
  \label{eq:boltzmann}
  \frac{\partial f}{\partial t} + v^i\frac{\partial f}{\partial x^i} +
  g^i\frac{\partial f}{\partial v^i} = 0,
\end{align}
also known as the collisionless Boltzmann equation.  In an expanding
Universe with scalefactor $a$, $t$ is the Newtonian time
($a^2dt=dt_p$), $x^i$ is the comoving coordinate ($adx^i=dx^i_p$),
$v^i$ is the conjugate velocity ($v^i=dx^i/dt$) and $g^i$ is the
gravitational acceleration field ($g^i=-a^2\partial\phi/\partial x^i$
where $\phi$ is the Newtonian potential)
\citep{bib:Bertschinger1995,bib:Inman2017a}.  The initial conditions
for the \cnb{} is the relativistic Fermi-Dirac distribution:
\begin{align}
  \label{eq:fermidirac}
  \bar{f}_\nu = \frac{\beta^3}{6 \pi \zeta (3) } \frac{1}{ \exp[\beta v]+1}
\end{align}
where $\beta=m_\nu/T_\nu$ and the normalization is chosen such that
the mean density is unity: $\bar{\rho}=\int d^3v \bar{f}(v) = 1$.

Eq.~\ref{eq:boltzmann} can be converted into a hierarchy of moment
equations.  We denote coarse-grained fields with the notation
$\langle A(t,x^i,v^i) \rangle = \int d^3v A f$.  We will concern
ourselves with the the density, $\rho = \langle 1 \rangle$, the
momentum $\rho V^i = \langle v^i \rangle$, the stress-energy tensor
$\rho \Sigma^{ij} = \langle v^iv^j \rangle$ and the third order heat
tensor $\rho Q^{ijk} = \langle v^i v^j v^k\rangle$.  The stress-energy
tensor is often expanded as
$\rho \Sigma^{ij} = \rho V^i V^j + \rho \sigma^{ij}$ where $V^i$ is
the velocity field and
$\sigma^{ij} = \rho^{-1}\langle (v^i-V^i)(v^j-V^j) \rangle$ is the
velocity dispersion. $\rho Q^{ijk}$ can be expanded as
$\rho Q^{ijk} = \rho V^iV^jV^k + \rho V^i \sigma^{jk} + \rho V^j
\sigma^{ik} + \rho V^k \sigma^{ij} + \rho q^{ijk}$
where $\rho q^{ijk} = \langle (v^i-V^i)(v^j-V^j)(v^k-V^k) \rangle$ is
heat flux tensor with associated heat flux $(1/2) \rho q^{iik}$.

By integrating Eq.~\ref{eq:boltzmann} over velocity we can obtain
their equations of motion.  These are the continuity equation:
\begin{align}
  \label{eq:continuity}
  \frac{\partial}{\partial t}\rho + \frac{\partial}{\partial x^k}
  \rho V^k= 0, 
\end{align}
the momentum equation:
\begin{align}
  \label{eq:momentum_s}
  \frac{\partial}{\partial t}\rho V^i + \frac{\partial}{\partial x^k} \rho
  \Sigma^{ik} = \rho g^i
\end{align}
and the stress-energy equation
\begin{align}
  \label{eq:stress_energy}
  \frac{\partial}{\partial t}\rho \Sigma^{ij} + \frac{\partial}{\partial x^k}\rho
  Q^{ijk} = \rho V^ig^j + \rho V^jg^i.
\end{align}

\subsection{Hydrodynamic Equations}
We wish to re-express these equations in terms of hydrodynamic
variables.  To that end, we expand the velocity dispersion in terms of
primitive variables $\rho \sigma^{ij} = P\delta^{ij} + \pi^{ij}$ where
$P$ is the pressure and $\pi^{ij}$ is the shear stress, which is
traceless, $\pi^{ii}=0$, and symmetric, $\pi^{ij}=\pi^{ji}$.  We now
require equations of motion for $P$ and $\pi^{ij}$ which can be
obtained from Eq.~\ref{eq:stress_energy}.  Instead of using the
stresses directly, it is convenient to use the energy, $E$, and
anisotropic stress, $\tau^{ij}$.  The energy is related to the trace
of the stress-energy tensor
\begin{align}
  E &= \frac{1}{2} \rho \Sigma^{ii} \nonumber \\
    &= \frac{1}{2}\rho V^2 + \frac{3}{2}P \nonumber \\
    &= \frac{1}{2}\rho V^2 +\frac{P}{\gamma-1}
\end{align}
which matches the definition of the energy of an ideal fluid with
$\gamma=5/3$ \citep{bib:Mitchell2013}.  The anisotropic stress is then
the traceless component:
\begin{align}
  \label{eq:anisotropic_stress}
  \tau^{ij} &= \rho \Sigma^{ij} - \frac{1}{3}\rho\Sigma^{kk}\delta^{ij}
  \\ \nonumber
            &= \rho V^iV^j-\frac{1}{3}\rho V^2 \delta^{ij} + \pi^{ij}.
\end{align}
The equations of motion in terms of the conservative variables are
then
\begin{align}
  &\frac{\partial}{\partial t}\rho V^i + \frac{\partial}{\partial
    x^k}\left[ \rho V^i V^k + P\delta^{ik} + \pi^{ik} \right] = \rho
    g^i, \label{eq:momentum} \\
  &\frac{\partial}{\partial t} E + \frac{\partial}{\partial
    x^k}\left[ EV^k+PV^k + V^l\pi^{lk} + \frac{1}{2}\rho q^{iik}
    \right] = \rho V^k g^k \label{eq:energy_q}
\end{align}
and

\begin{align}
  \label{eq:anisotropic_q}
  \frac{\partial}{\partial t} \tau^{ij} &+
                                          \frac{\partial}{\partial x^k} \left[ \tau^{ij}V^k + 
                                          PV^i\delta^{jk} +
                                          PV^j\delta^{ik} -\frac{2}{3} P V^k\delta^{ij} \right. \nonumber\\
                                        &\left.  + V^i\pi^{jk}
                                          + V^j \pi^{ik} - \frac{2}{3}
                                          V^l\pi^{lk}\delta^{ij}
                                          + \rho
                                          q^{ijk} - \frac{1}{3} \rho
                                          q^{llk}\delta^{ij} \right] \nonumber\\
                                        &=
                                          \rho V^i g^j + \rho V^j g^i - \frac{2}{3} \rho V^k g^k \delta^{ij}
\end{align}

\subsection{Closing the Hierarchy}
Solving these equations is exceptionally challenging due to the
non-trivial stresses and heat tensor which are present even in linear
theory.  Strategies that have been employed to estimate the higher
moments include linear theory \citep{bib:Dakin2017}, using N-body
particles to estimate higher moments
\citep{bib:Pueblas2009,bib:Bannerjee2016}, or the ``Zero Heat Flux''
approximation which sets $\rho q^{ijk}=0$
\citep{bib:Vorobyov2006,bib:Mitchell2013}.

An alternate approach is to try to deduce a (potentially nonlinear)
equation of state for the heat flux which depends only on the
hydrodynamic variables (see,
e.g.,~\citet{bib:Hammett1990,bib:Chust2006,bib:Wang2015}).  To do this
we will consider two test problems: the free evolution of particles
and the linearized evolution of phase space.
    
\subsubsection{Free Evolution of Particles}
We first consider the free evolution ($g^i=0$) of collisionless
particles.  In this case, the exact solution to Eq.~\ref{eq:boltzmann}
is given in Fourier space by:
\begin{align}
  \ft[f]=\ft[f(t_i)] \exp[-i k^j v^j (t-t_i)].
\end{align}
If we now consider an initial distribution consisting solely of
particles with the same velocity:
$f(t_i) \propto \rho(t_i,x^i) / (4 \pi v^2) \delta_D(v-u)$ the density
field is given by:
\begin{align}
  \label{eq:isl}
  \ft[\rho] = \ft[\rho(t_i,x^i)] j_0(k u (t-t_i)).
\end{align}
The inverse Fourier transform of $j_0(k u (t-t_i))$ is
$\delta_D(r-u(t-t_i))/(4\pi r^2)$ and so this can be thought of as the
convolution of the initial density field with the inverse square law.
In general, Eq.~\ref{eq:isl} satisfies the wave-like spherical Bessel
equation in real space:
\begin{align}
  \frac{\partial^2}{\partial t^2} \rho + \frac{2}{t-t_i}
  \frac{\partial}{\partial t} \rho - u^2
  \frac{\partial^2}{\partial x^i \partial x^i} \rho = 0.
\end{align}
The \cnb{} is initially homogeneous, $\rho(t_i,x^i)=1$, and so no
direct convolution needs to be performed.

\subsubsection{Linearized Vlasov Equation}
We next consider the Vlasov equation linearized about some initial
homogeneous and isotropic velocity distribution
$f(t_i,x,v)=\bar{f}(v) $:
\begin{align}
  \frac{\partial f}{\partial t} + v^i\frac{\partial f}{\partial x^i} +
  g^i\frac{\partial \bar{f}}{\partial v^i} = 0.
\end{align}
We first remark that ``linearized'' is a misnomer in our particular
application: since neutrinos are a small component ($f_\nu\ll1$) the
Vlasov equation is already approximately linear as $g^i$ is
approximately independent of neutrino density.  Instead, we are
replacing a derivative term with a source term.  This is precisely why
the ``linear response'' method does not produce accurate neutrino
power spectra despite neutrinos remaining linear
(e.g.~$\delta_\nu \ll 1$) at all times.  Nonetheless, we will use the
standard terminology throughout.  The linear response solution to this
equation is given in Fourier space by
\citep{bib:Gilbert1966,bib:AliHaimoud2013}:
\begin{align}
  \ft[f] = \int_{t_i}^t dt'(-\ft[g^i]) \frac{\partial
  \bar{f}}{\partial v^i} \exp[-ik^jv^j(t-t')].
\end{align}
The moments are then given by:
\begin{widetext}
  \begin{align}
    \ft[\rho] &= \int_0^\infty dv 4 \pi v^2 \bar{f}(v)\int
                _{t_i}^{t} dt' (-ik^j\ft[g^j])(t-t') j_0(kv(t-t')) \label{eq:linearsoln}\\
    ik^i\ft[\rho V^i] &= \int_0^\infty dv 4 \pi v^2 \bar{f}(v) 
                        \int_{t_i}^t dt' (ik^j\ft[g^j]\cos[kv(t-t')] \\
    ik^iik^j\ft[\rho\Sigma^{ij}] &=\int_0^\infty dv 4 \pi v^2
                                   \bar{f}(v) v^2k^2 \int
                                   _{t_i}^{t} dt' (-ik^j\ft[g^j])(t-t')
                                   j_0(kv(t-t')) \\
    ik^iik^jik^k\ft[\rho Q^{ijk}] &=\int_0^\infty dv 4 \pi v^2
                                    \bar{f}(v) v^2k^2 \int_{t_i}^t dt' (ik^j\ft[g^j]\cos[kv(t-t')]
  \end{align}
\end{widetext}
where we have omitted the homogeneous terms.  It is immediately clear
that density fields can be composed as sums over velocity shells.
That is, if the background distribution with
$\bar{f}_u = 1/(4\pi v^2)\delta_D(v-u)$ has moments such as $\rho_u$,
then a more general distribution function $\bar{f}$ has
$\rho = \int_0^\infty du 4\pi u^2 \bar{f} \rho_u$
\citep{bib:Inman2017a}.  Furthermore, these shells have simple higher
moments.  In particular, the stress-energy tensor can be written
entirely as an isotropic pressure
$\rho \Sigma^{ij} = \rho \sigma^{ij} = P\delta^{ij} = \rho u^2
\delta^{ij}$ and so the density satisfies the wave equation:
\begin{align}
  \label{eq:wave}
  \frac{\partial^2}{\partial t^2} \rho - u^2
  \frac{\partial^2}{\partial x^i \partial x^i} \rho =
  -\frac{\partial}{\partial x^i} g^i.
\end{align}
Substituting this relationship and the linearized
Eq.~\ref{eq:continuity} into the linearized Eq.~\ref{eq:stress_energy}
leads to:
\begin{align}
  \frac{\partial}{\partial x^k} Q^{ijk} &=
                                          \frac{\partial}{\partial x^k}(u^2 V^k \delta^{ij})
                                          \nonumber \\
                                        &=
                                          \frac{\partial}{\partial
                                          x^k}\left[ u^2 (V^i
                                          \delta^{jk} +
                                          V^j\delta^{ik} +
                                          V^k\delta^{ij} ) +
                                          q^{ijk}\right]
\end{align}
or
\begin{align}
  \label{eq:general_grad_q}
  \frac{\partial}{\partial x^k}  q^{ijk} &=
                                           \frac{\partial}{\partial x^k} \left[ -  u^2 V^i
                                           \delta^{jk} - u^2 V^j
                                           \delta^{ik} \right]
                                           \nonumber \\
                                         &=-u^2\frac{\partial
                                           V^i}{\partial x^j} -
                                           u^2 \frac{\partial
                                           V^j}{\partial x^i}.
\end{align}
Linear velocity fields are curl free and can be written in terms of a
velocity potential $V^i = \partial \phi_v/\partial x^i$ which leads to
\begin{align}
  \frac{\partial}{\partial x^k}  q^{ijk} &=
                                           -2 u^2 \frac{\partial^2
                                           \phi_v}{\partial
                                           x^i \partial x^j}.
\end{align}
The symmetric form of $q^{ijk}$ satisfying this is then given by:
\begin{align}
  \label{eq:general_q}
  q^{ijk} &=-2 u^2 \frac{\partial}{\partial x^i}
            \frac{\partial}{\partial x^j} \frac{\partial}{\partial
            x^k} \nabla^{-2} \phi_v\\
          &=-\frac{2}{3}u^2 \nabla^{-2} \left[
            \frac{\partial}{\partial x^i}\frac{\partial}{\partial x^j}V^k +
            \frac{\partial}{\partial x^i}\frac{\partial}{\partial x^k}V^j +
            \frac{\partial}{\partial x^j}\frac{\partial}{\partial x^k}V^i \right]
\end{align}
where $\nabla^{-2}$ denotes the inverse Laplacian.
       
\subsubsection{Shell Equations of Motion}
Unfortunately, it is unclear how to best generalize the linear heat
flux of a shell once nonlinear evolution occurs.  Simply using
Eq.~\ref{eq:general_grad_q} for the heat tensor leads to an energy
flux of $(P-\bar{\rho} u^2) V^k+\pi^{ij}$ which could be $0$, the
nonlinear fluctuation of $P$ or something else entirely.  Lacking a
nonlinear scheme, we instead opt for a particularly simple closure
choice that allows for the shell equations to be independent of $u$:
\begin{align}
  \label{eq:qsimple}
  \frac{\partial}{\partial x^k}\rho q^{ijk} =
  \frac{\partial}{\partial x^k} \left( -\rho\sigma^{ik}V^j -
  \rho\sigma^{jk}V^i \right)
\end{align}
which yields the much simplified continuity equations for
stress-energy:
\begin{align}
  \label{eq:Esimple}
  \frac{\partial }{\partial t} E + \frac{\partial}{\partial x^k}
  EV^k &= \rho V^k g^k \\
  \label{eq:tausimple}
  \frac{\partial }{\partial t} \tau^{ij} + \frac{\partial}{\partial x^k}
  \tau^{ij}V^k &= \rho V^i g^j + \rho V^jg^i - \frac{2}{3} \rho V^k
                 g^k \delta^{ij}.
\end{align}
The complete set of equations for a shell is therefore:
Eqs.~\ref{eq:continuity}, \ref{eq:momentum}, \ref{eq:Esimple} and
\ref{eq:tausimple}.  In the absence of gravity, these equations
collapse down to wave dynamics.  We note that the wave equation allows
for superposition which resembles shell crossing.  In this work we
make one further simplification and neglect the shear stress entirely,
$\pi^{ij}=0$, leading to {\it ideal} collisionless hydrodynamics.
This simplification is still consistent with linear theory but allows
us to save substantial resources by not solving
Eq.~\ref{eq:tausimple}.  We consider two alternative schemes, the
isothermal and ideal gas approximations, in the Appendix.

Let us lastly comment on the difference between this approach and
linear response.  In linear response approximations are made to the
gravitational terms, i.e.~$\rho g^i \rightarrow \bar{\rho}g^i$.  In
Eqs.~\ref{eq:Esimple} and \ref{eq:tausimple} we have made
approximations to the hydrodynamic fluxes, but kept gravitational
terms fully intact.  We therefore expect this approach to work
significantly better as it retains linearity in the neutrino
perturbations themselves.

\section{Methods}
\label{sec:methods}
In this section we describe our implementation of the neutrino solver
into the \cube{} code \citep{bib:Yu2018}.  \cube{} is a refactored
version of \cpm{} \citep{bib:HarnoisDeraps2013} designed to use the
minimum amount of memory possible: 6 bytes per particle.  It features
a three level force decomposition: a ``coarse'' particle mesh force
over $n_c$ cells where the only global FFT is performed, a ``fine''
particle mesh force with resolution $n_f=4 n_c$ where local FFTs are
performed, and a direct particle-particle (PP) force at the cell
level.  \cube{} is parallelized with both OpenMP and co-array Fortran.
Both \cube{} and \cpm{} are publicly available.

\subsection{Single-Shell Hydrodynamics}
\label{ssec:hydro}
In principle, any method to solve the hydrodynamical equations can be
used for collisionless hydrodynamics.  For our purposes, we have opted
for a simple dimensionally split Eulerian grid method.  Overall, the
code is very similar to the example code of \citet{bib:Trac2003},
except with the relaxing algorithm replaced by an HLL approximate
Riemann solver \citep{bib:Harten1983}.  While we don't expect
neutrinos to require particularly high spatial resolution, we allow
for it by using a piecewise linear flux calculation with a nonlinear
flux limiter.  The gravitational force is considered as a source term,
and coupled to the N-body code in the same way as the
magnetohydrodynamics code was coupled to \cpm{}.  This means that
density is conserved to machine precision, but momentum and energy are
not.  Since the hydrodynamic grid need not have the same number of
cells as the gravitational grid, we simply interpolate the values of
all gravitational forces computed within a hydrodynamical cell.  If
the grid is coarser than $n_f/2$ or $n_c/2$ this introduces force
artifacts on the fine/coarse grid scale since only one cell is ever
buffered.  Since this is, by construction, subgrid for the neutrinos,
and indeed the neutrino perturbations are themselves highly suppressed
anyways, we have not encountered any issues.

\subsection{Cosmic Neutrino Background}
\label{ssec:cnbhydro}
Utilizing this technique for the the \cnb{} requires the simulation of
multiple velocity shells, each with density $\rho_u$, which can then
be integrated over to obtain the neutrino density:
$\rho_\nu = \int_0^\infty du 4 \pi u^2 \bar{f}_\nu \rho_u$.  The shell
densities can be computed using the methods of
Subsection~\ref{ssec:hydro}.  Optimally performing this integration
was studied by \citet{bib:Lesgourgues2011} who found that
Gauss-Laguerre integration is the best choice.  We therefore use this
quadrature strategy and find that only three integration points
provide accurate results.  The proper shell velocities are then
$\{418,2310,6320\} \times (1+z) (0.05\ev/m_\nu) $ \kms{} with the
fastest shell becoming non-relativistic at $z\sim 50 (m_\nu/0.05\ev)$
\kms.

Since CDM perturbations are typically set at $z\sim100$ and we do not
want to be evolving neutrinos until they are safely non-relativistic,
we employ the ``late start'' approach developed in
\citet{bib:Inman2015}.  Specifically, CDM perturbations are generated
at some intermediate redshift $z_\nu$ and then propagated backwards to
the initial redshift $z_i$ with a growth factor that assumes neutrinos
are homogeneous: $D_+\sim (a_i/a_\nu)^{1-3f_\nu/5}$
\citep{bib:Bond1980}.  The forward evolution proceeds in two phases:
between $z_i$ and $z_\nu$ the CDM evolves alone (i.e.~$\delta_\nu=0$).
This precisely undoes the modified growth factor on linear scales
whereas on nonlinear scales the neutrino perturbations are negligible
anyways.  After $z_\nu$ both the CDM and the \cnb{} are simultaneously
evolved.

Since the shell velocities are quite large, the neutrino timestep is
required to be quite small.  We allow each velocity shell to have its
own grid size which allows faster shells to be coarser and hence have
shorter timesteps.  A future optimization is to allow neutrinos to
take multiple timesteps between each CDM one.  It may also be possible
to allow shells to start at different redshifts, e.g., when their
perturbations become dynamically relevant for the simulation.

\subsection{Initial Conditions}
Since we do not have transfer functions for individual momenta
(although in principle these can be extracted from Boltzmann codes),
we opt for approximate initial conditions for each velocity shell.  To
determine an approximate solution for each shell, we first consider
their linear equations of motion, Eq.~\ref{eq:wave} (this can also be
seen by differentiating Eq.~\ref{eq:linearsoln} twice):
\begin{align}
  \frac{\partial^2 \delta_u}{\partial t^2} + k^2 u^2 \delta_u = a^2 (-k^2\phi).
\end{align}
On large scales, where $k$ is small, the second term is negligible and
$\delta_u\simeq\delta_c$.  On small scales, the fluid responds
instantaneously and the solution is
$\delta_u\simeq (k/k_{u})^2\delta_c$ where
$k_{u}=\sqrt{(3/2)\Omega_m a}H_0/u$ is the free streaming scale of the
shell \citep{bib:Inman2017a}.  We utilize the following approximation:
\begin{align}
  \rho_u = 1 + \delta_u =1 + \delta_c\frac{1}{1+\lambda k/k_u + (k/k_u)^2}.
\end{align} 
The corresponding velocity field is obtained via the continuity
equation:
\begin{align}
  V^i_u = V^i_c\frac{1+1.5 \lambda k/k_u + 2(k/k_u)^2}{(1+\lambda k/k_u + (k/k_u)^2)^2}.
\end{align}
In the simulations used in this paper we used $\lambda=1.$ which is
appropriate for $z_\nu=0$; however, we have subsequently solved for
the optimal value of $\lambda$ and find that $\lambda=0.75,0.76,0.77,$
and $0.82$ for $m_\nu=0.05,0.10,0.20$ and $0.40$ at $z_\nu=10$.  These
values are are rather unchanged at $z_\nu=5$ but increase towards $1$
at $z_\nu=0$.  Utilizing a constant $\lambda=0.75$ at either
$z_\nu=10$ or $5$ produces neutrino density and velocity transfer
functions accurate to around 10\% regardless of neutrino mass.
    
When computing the initial momentum density we opt to include the
second order term, i.e.~$\rho V^k = (1+\delta_u) V_u^k$ rather than
just $V_u^k$.  Likewise, the initial energy is therefore
\begin{align}
  E_u=\rho_u\left[ \frac{1}{2} V_u^2 + \frac{u^2}{\gamma-1} \right]
\end{align}
with the kinetic term included.

\section{Results}
\label{sec:results}
We run simulations of neutrinos of masses 0.05 \ev, 0.10 \ev, 0.20
\ev{} and 0.40 \ev{} using both \cube{} and \cpm.  We keep the initial
Gaussian noise realization equivalent between the two simulations and
use the same number of particles, but otherwise allow them to run
independently.  Because the neutrinos are not coupled to the
particle-particle force, we run \cube{} without the direct force.
Importantly, this means that the \cpm{} simulations have better force
resolution as they include the pairwise force on small scales.  For
cosmological parameters, we match those of \citet{bib:Cataneo2019} so
that we can compare to the high resolution results computed in that
work.  Specifically, we use a flat universe with fixed
$1-\Omega_\Lambda=\Omega_m=\Omega_c+\Omega_b+\Omega_\nu=0.291$,
$\Omega_b=0.047$, $h=0.6898$, and implicitly varying $\Omega_c$ via
$\Omega_\nu=m_\nu/93.14/h^2$.  For initial conditions we use
$A_s=2.442\times 10^{-9}$ and $n_s=0.969$.  In all cases we use
$256^3$ CDM particles and simulate only a single massive neutrino
species.  In \cpm{} the \cnb{} is resolved by $256^3$ N-body
particles, whereas in \cube{} they are resolved by three fluids with
grid resolution of $256^3$.  The only exception is for the
$m_\nu=0.05$ \ev{} case where we used $128^3$ grid cells for the
fastest shell.  In this simulation we also decreased the CFL condition
from 0.7 to 0.5 as we noticed some slight oscillations on small scales
in the output.  In all simulations for both codes we use $z_i=100$ and
$z_\nu=10$.  We show slices of CDM and \cnb{} density fields in
Fig.~\ref{fig:slices} which demonstrate well the lack of Poisson noise
in the hydrodynamic simulations.

\begin{figure}
  \begin{center}
    \includegraphics[width=0.45\textwidth]{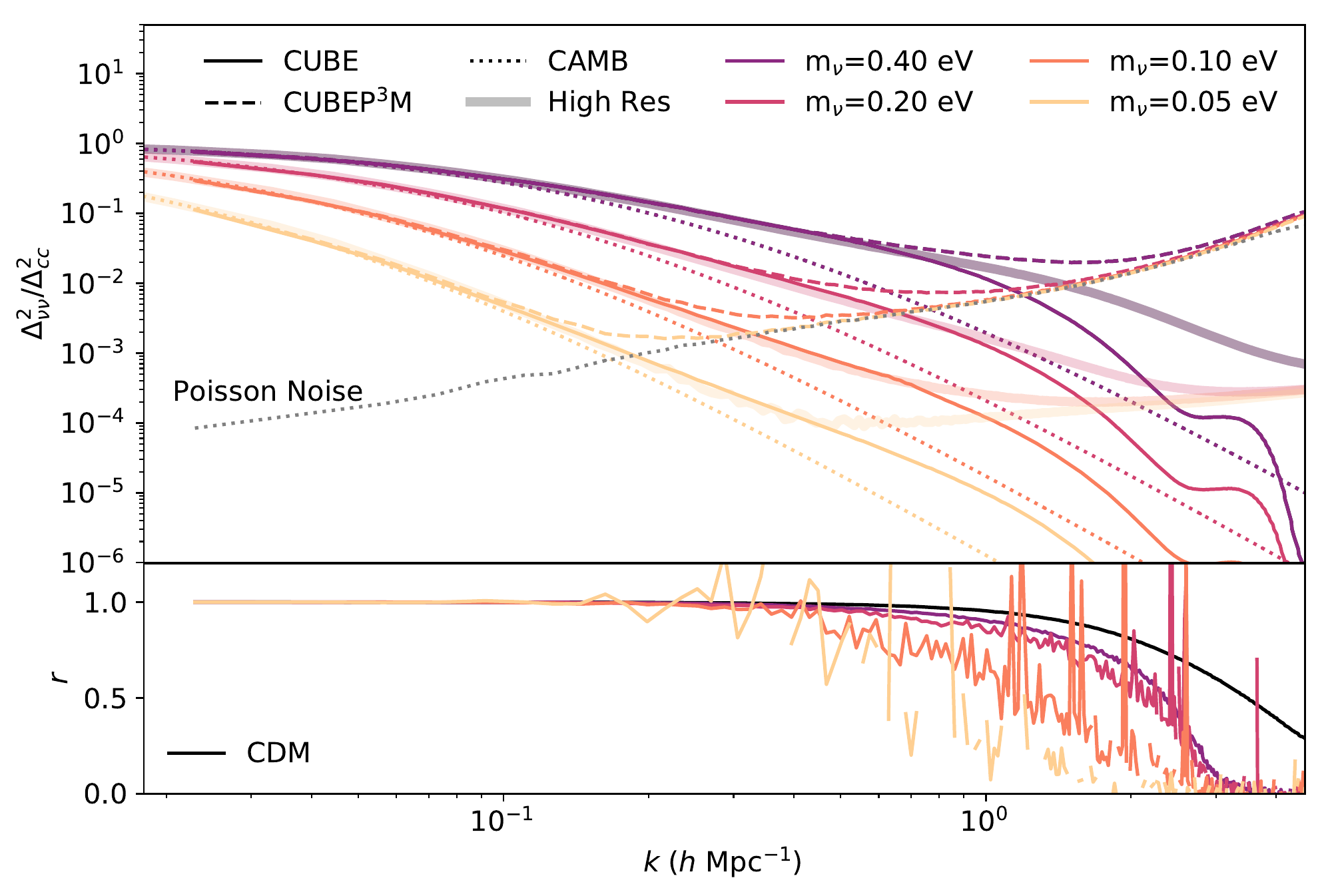}
    \caption{(Top) Ratio of \cnb{} power spectra to the CDM one from
      hydrodynamics (solid), N-body (dashed) and linear theory
      (dotted).  Also shown are the theoretical Poisson noise (dotted
      grey) and results from high resolution N-body simulations from
      \citet{bib:Cataneo2019} (solid bands).  (Bottom) The correlation
      coefficient between \cube{} and \cpm{}.  The black band is the
      coefficient between CDM in the neutrinoless simulations.}
    \label{fig:powerM}
  \end{center}
\end{figure}

We show the ratio of the \cnb{} power spectra to the CDM one in the
top panel of Fig.~\ref{fig:powerM}.  Solid lines utilize the
collisionless hydrodynamics approach, dashed lines the particles,
dotted lines are linear theory results computed using CAMB
\citep{bib:Lewis2000}, and bands are the high-resolution simulations
from \citet{bib:Cataneo2019}\footnote{We note that there is still
  noise in these power spectra.  This may be due to the random number
  generator utilized by the compiler of the supercomputer these
  simulations were run on, as it does not occur on other machines.}.
For the lighter neutrino species the hydrodynamic simulations provide
significantly better resolution due to lack of Poisson noise, whereas
for the heavier ones it is merely comparable (although a lack of power
seems preferable to an artificial enhancement).  In the bottom panel
we show the correlation coefficient between the two codes, defined as
the ratio of the crosspower spectrum to the square root of the
geometric mean of the autopower spectra.  In general, this would go to
zero when Poisson noise kicks in.  To cancel shot noise in the power
spectrum, we divide the neutrino particles randomly into two groups
and compute the cross power as described in \citet{bib:Inman2015}.
For $m_\nu < 0.2$ \ev{} this is very noisy due to the low particle
number.  However, for the heavier neutrino species it does do quite
well and we find the hydrodynamics approach is well correlated to at
least $k\sim 1$ \mpch.

\begin{figure}
  \begin{center}
    \includegraphics[width=0.45\textwidth]{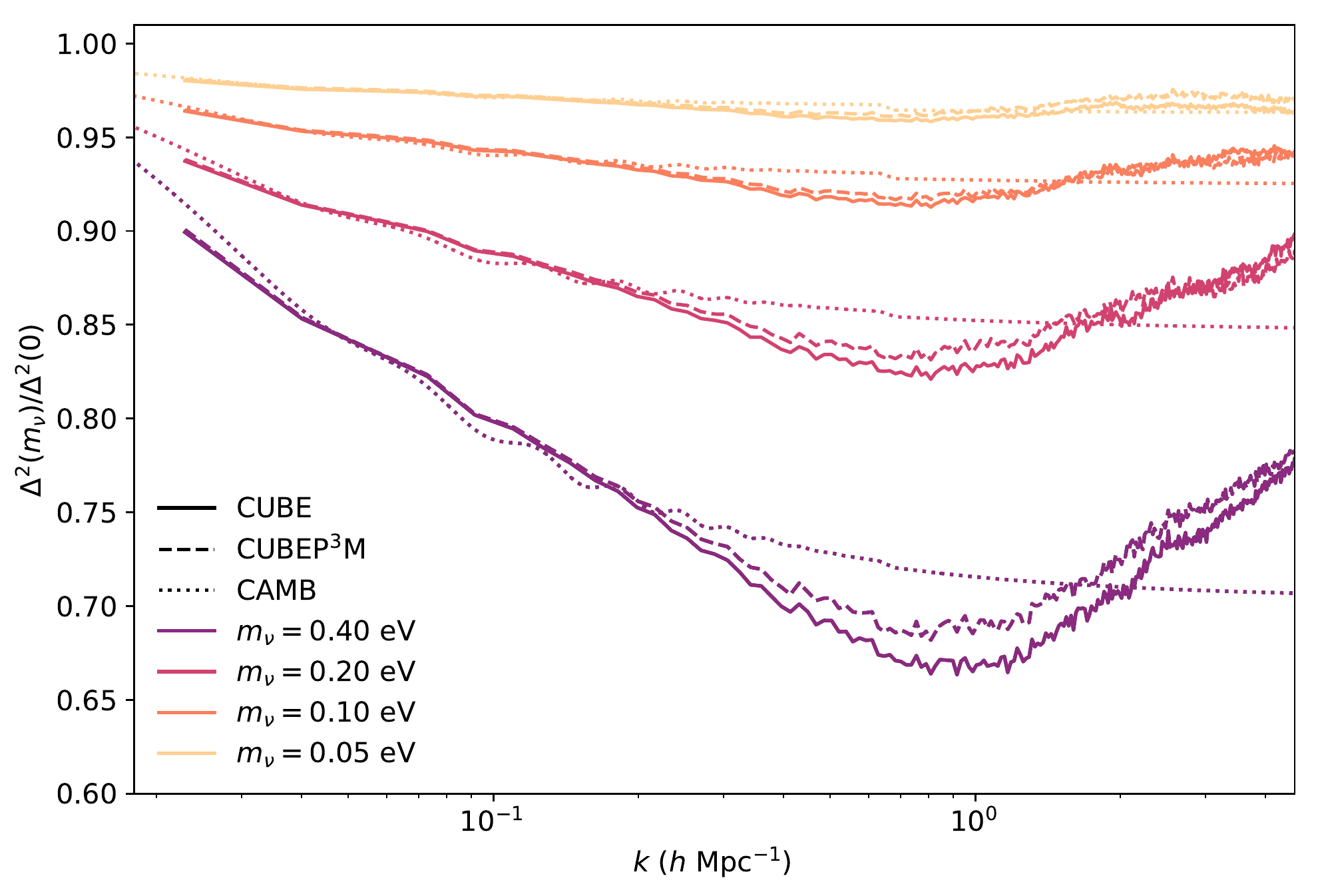}
    \caption{The suppression of the matter power spectrum as a
      function of neutrino mass.  Dotted lines are linear theory from
      \camb, dashed are computed using \cpm, and solid are from
      \cube.}
    \label{fig:ratioM}
  \end{center}
\end{figure}

The modulation of the matter power spectrum is shown in
Fig.~\ref{fig:ratioM}.  \cube{} yields the same result as \cpm,
matching linear theory on large scales but with an enhanced dip on
nonlinear ones.  On nonlinear scales \cube{} and \cpm{} very slightly
disagree.  We suspect that this is due to the difference in force
calculation between the codes which results in slightly different
power spectra.

\section{Discussion}
\label{sec:discussion}
In an ideal simulation method neutrinos would increase memory and
computation time by $\mathcal{O}(f_\nu)$.  Our method can satisfy the
first criterion easily.  As an example, we consider the hypothetical
requirements to run a TianNu scale simulation \citep{bib:Emberson2017}
with the \cube{} code.  TianNu evolved nearly three trillion
particles, mostly neutrinos, in a cubic volume of width 1200 \mpch.
Due to Poisson noise, TianNu only resolved neutrino perturbations to
$k\sim 1$ \hmpc{} which generically requires a grid with $\sim 1024$
cells per dimension to resolve.  With our new hydrodynamical method,
this would require $3\times5\times1024^3$ floating point numbers.
This may be compared to the number required in TianNu,
$6\times13824^3$ floating point numbers, a savings of approximately
$\sim10^3$.  The memory usage could be further decreased if faster
shells are allowed to be less resolved.  \cube{} also has a
significant CDM memory compression scheme which allows CDM memory
reduction to as low as 6 bytes per particle (from 28 in \cpm).  In
total, the TianNu simulation could be run utilizing around 40 times
less memory to store particles.

\begin{figure}
  \begin{center}
    \includegraphics[width=0.45\textwidth]{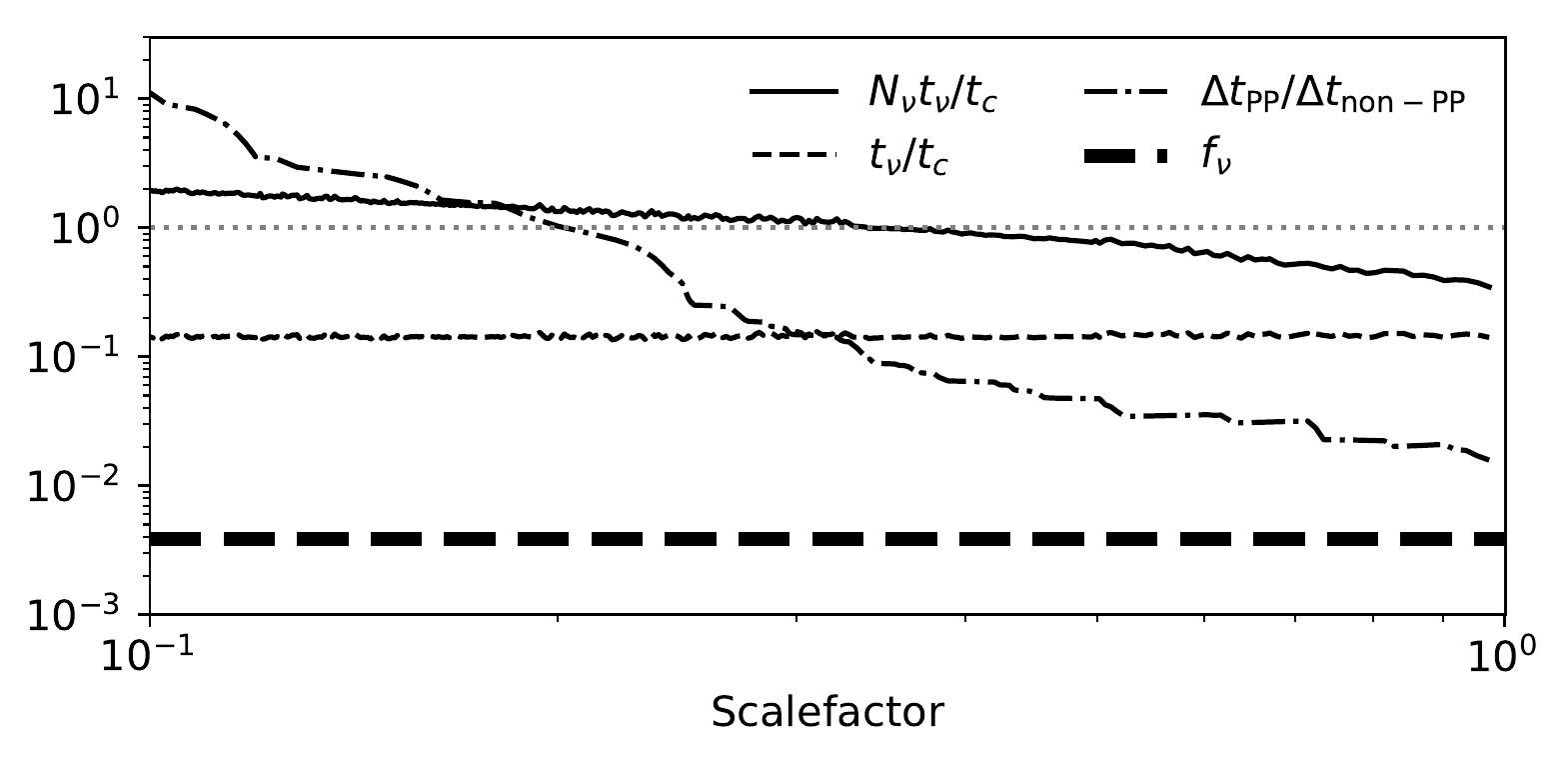}
    \caption{Relative time taken in neutrino and CDM computations from
      the $0.05$ \ev{} simulation.  The dashed line is the ratio of
      the physical time per timestep spent in the neutrino
      hydrodynamic subroutines to the CDM subroutines.  The solid
      black line is the hypothetical ratio if individual timesteps are
      used for the neutrinos.  The dash-dotted line is the ratio of
      the PP timestep to non-PP timesteps in a CDM only simulation.}
    \label{fig:timing}
  \end{center}
\end{figure}

In our current implementation the neutrino computations are quite fast
compared to the CDM ones, however the total computation time is
extended well beyond an $\mathcal{O}(f_\nu)$ increase due to the short
time step required by the neutrinos.  Nonetheless, we are optimistic
that this can be resolved by individual timestepping, where neutrinos
take many more short timesteps than CDM.  To obtain an idea of how
much improvement can be obtained, we extract various timing quantities
at each timestep: the time taken in the CDM subroutines ($t_c$ - note
that this includes the force calculation, which has some neutrino
operations as well), the time taken in the neutrino hydrodynamic
routines ($t_\nu$), the required CDM timestep ($\Delta t_c$) and the
required neutrino timestep ($\Delta t_\nu$) utilizing CFL=0.5.  The
number of neutrino steps that can be taken per CDM one is then given
by $N_\nu=\Delta t_c/\Delta t_\nu$.

In Fig.~\ref{fig:timing} we show the ratio of the time necessary for
neutrino calculations $N_\nu t_\nu$ compared to the time needed for
CDM ones $t_c$ for the most challenging $0.05$ \ev{} case.  At high
redshifts, the neutrino computations are substantial; however, at
lower redshifts they quickly become subdominant.  \citet{bib:Bird2018}
demonstrated that neutrinos with velocities $\gtrsim 100$ \kms{} can
be treated with linear response down to $z\sim1$ and so we expect that
we can always set $z_\nu$ such that neutrinos are subdominant.  Indeed
when the subgrid pairwise force, which requires a much smaller
timestep as well, is also used the neutrinos may even become a
negligible fraction.  We ran a single CDM only simulation using
\cube{} with the PP force enabled with an extended range of one fine
cell.  The dash-dotted line in Fig.~\ref{fig:timing} shows the ratio
of the timestep required for the PP force compared to other non-PP
timesteps, giving an estimate of the scaling when subgrid forces are
included.  \cube{} is currently undergoing optimizations for upcoming
large scale simulations, and we hope to include individual
timestepping in a subsequent update.

\section{Conclusion}
\label{sec:conclusion}
We have solved the Vlasov equation for the \cnb{} using a novel
technique based on decomposing the homogeneous phase space into
velocity shells which are then evolved via a closed set of Boltzmann
moment equations.  We have found this approach to accurately model the
neutrino perturbations well into the nonlinear regime.  This method
does not have Poisson noise, which plagues the more common particle
based methods.

To model the \cnb{} we have been interested in subsonic shells
($V \ll u$).  A natural extension is to look for approximate closure
schemes in the supersonic regime, which could be useful for warm or
cold dark matter.  When simulated with particles warm dark matter is
known to exhibit artificial fragmentation \citep{bib:Wang2007} which
may be ameliorated through the use of hydrodynamic equations.  Of
course, while the equations we employed are consistent with CDM linear
evolution\footnote{Note that taking $u\rightarrow0$ in
  Eq.~\ref{eq:general_grad_q} yields the zero heat flux equations,
  which differ from Eqs.~\ref{eq:Esimple} and \ref{eq:tausimple}.},
and do allow for some level of shell crossing, it would be quite
shocking if they provided a useful model of CDM.  It is intriguing,
however, to consider the closure relation for a known system, the
Schr\"{o}dinger equation, which approximates CDM above the de Broglie
wavelength.  For the coarse-grained Wigner approximation, the exact
solution is given by
$q^{ijk} = \hbar^2/(4 m^2) \nabla^i \nabla^j \nabla^k \phi_v$
\citep{bib:Uhlemann2014}, which is very analogous to
Eq.~\ref{eq:general_q} with the ``momentum'' transformed to a quantum
operator: $\sqrt{8} m u \rightarrow -i \hbar \nabla$.  This is simply
due to the fact that linear theory and the Schr\"{o}dinger method both
have velocity potentials, but nonetheless the similar form of the heat
flux could provide a starting point for further investigations.  It
may also be more natural to treat potential dark matter interactions
with hydrodynamic equations (e.g.~\citet{bib:Kummer2019}).

\acknowledgements{ We are extremely grateful to Andrew MacFadyen for
  providing notes on implementing hydrodynamics.  We thank Yacine
  Ali-Ha\"{i}moud for valuable discussions and JD Emberson for
  assistance with the power spectra of \citet{bib:Cataneo2019}.  HRY
  acknowledges National Science Foundation of China No.11903021.  This
  work was supported in part through the NYU IT High Performance
  Computing resources, services and staff expertise.  This research
  has made use of NASA's Astrophysics Data System Bibliographic
  Services.}

\software{ {\sc NumPy} \citep{bib:Walt2011}, {\sc Matplotlib}
  \citep{bib:Hunter2007}, {\sc SciPy} \citep{bib:Jones2001} }

\appendix

\section{Isothermal and Ideal Gas Approximations}
\label{app:iga}
\begin{figure}
  \begin{center}
    \includegraphics[width=0.45\textwidth]{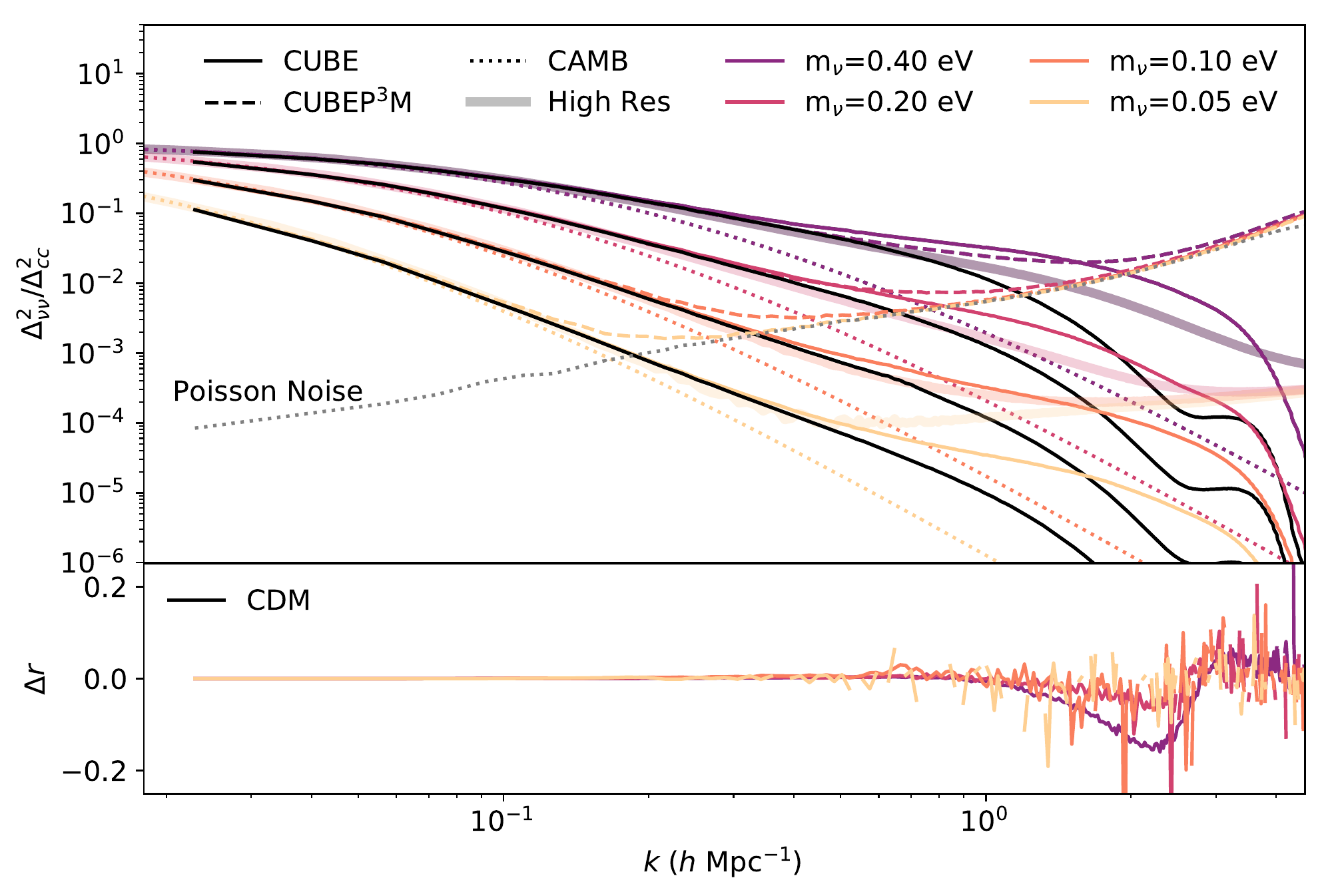}
    \caption{(Top) Ratio of \cnb{} power spectra, calculated using the
      isothermal gas approximation, to the CDM one.  Black curves are
      the results from Fig.~\ref{fig:powerM}.  (Bottom) Difference in
      correlation coefficient between iosthermal gas and N-body, and
      collisionless hydrodynamics and N-body.}
    \label{fig:powerIS}
  \end{center}
\end{figure}

In this section we test to what extent the isothermal and ideal gas
approximations can be used to model the shells.  For the isothermal
gas, we assume that $P=\rho u^2$ at all times, with soundspeed
$c_s=u$.  We show the results in Fig.~\ref{fig:powerIS}.  In the top
panel the solid coloured lines show the isothermal approximation,
whereas the solid black lines show the collisionless hydrodynamics
result.  The bottom panel shows the difference in correlation
coefficient between the isothermal gas and the collisionless
hydrodynamics, i.e.~negative values indicate collisionless
hydrodynamics is better correlated.  We find that the isothermal
approximation appears to overpredict neutrino perturbations, but
remains well correlated with the particle field.

\begin{figure}
  \begin{center}
    \includegraphics[width=0.45\textwidth]{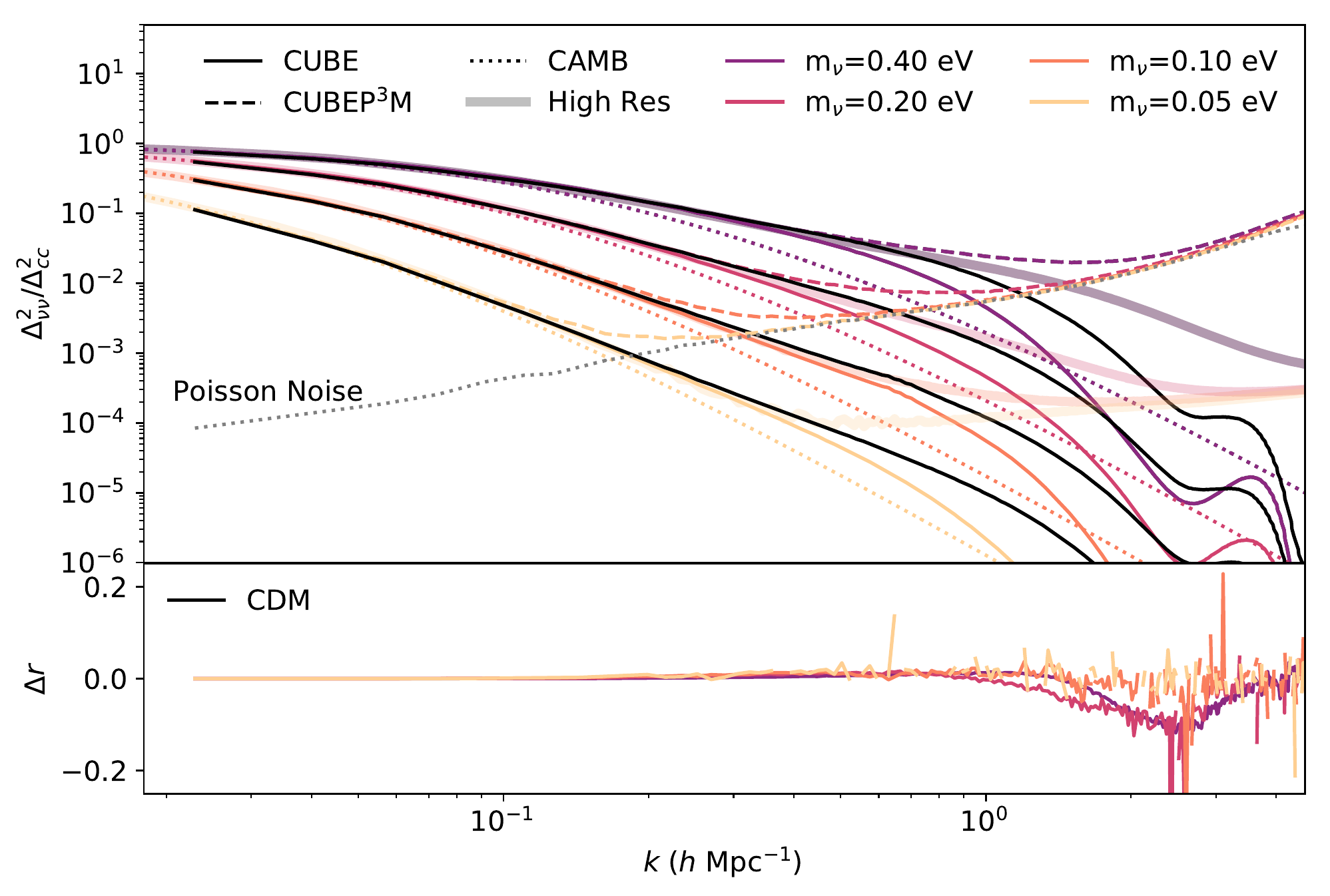}
    \caption{Same as Fig.~\ref{fig:powerIG} but using the ideal gas
      approximation.}
    \label{fig:powerIG}
  \end{center}
\end{figure}

For the ideal gas approximation, we employ the ideal gas energy
equation:
\begin{align}
  \frac{\partial}{\partial t} E + \frac{\partial}{\partial x^k}
  \left[ (E+P)V^k \right] = \rho V^k g^k.
\end{align}
Because the flux $(E+P)V^k$ is simply proportional to $V^k$ upon
linearization, the same as in collisionless hydrodynamics, we expect
that we can reproduce linear evolution by simply changing the initial
energy to:
\begin{align}
  E_u=\rho_u\left[ \frac{1}{2} V_u^2 + \frac{u^2}{\gamma(\gamma-1)} \right]
\end{align}
and the sound speed is the standard $\sqrt{\gamma P/\rho}$.  We show
the result in Fig.~\ref{fig:powerIG} which shows that the ideal gas
approximation slightly underpredicts neutrino perturbations.  Again,
this does not result in a substantially worsened correlation
coefficient.

We conclude that both approximations are sufficient to model the \cnb,
although perhaps slightly less accurate than collisionless
hydrodynamics.  As both isothermal and ideal gases are commonly
included in cosmological codes it is therefore straightforward for
these codes to include Poisson-noise free neutrino perturbations
provided they can simulate multiple fluids simultaneously.

\bibliographystyle{apsrev}
\bibliography{thebib}

\end{document}